\documentclass{aa}
\usepackage{psfig}
\usepackage{epsfig}

\def \sax {BeppoSAX}
\def \src {G0.9+0.1}

\def \deg{^\circ}

\def \hcm {\hbox {\ifmmode $ cm$^{-2}\else cm$^{-2}$\fi}}
\def \arcmin {\hbox{$^\prime$}}

\def\approxgt{\mathrel{\hbox{\rlap{\lower.55ex \hbox {$\sim$}}
        \kern-.3em \raise.4ex \hbox{$>$}}}}
\def\approxlt{\mathrel{\hbox{\rlap{\lower.55ex \hbox {$\sim$}}
        \kern-.3em \raise.4ex \hbox{$<$}}}}

\begin{document}

\thesaurus{08.(09.19.2; 09.09.1; 13.25.4)}

\title{X--ray emission from the galactic center region supernova remnant G0.9+0.1}

\author{L. Sidoli\inst{1} 
        \and S. Mereghetti\inst{2}
        \and G.L. Israel\inst{3,4} 
        \and F. Bocchino\inst{1}  
}
\offprints{L. Sidoli (lsidoli@astro.estec.esa.nl)}

\institute{
       Astrophysics Division, Space Science Department of ESA, ESTEC,
       Postbus 299, NL-2200 AG Noordwijk, The Netherlands
\and
       Istituto di Fisica Cosmica G.Occhialini,  C.N.R., Via Bassini 15, I-20133 Milano,
       Italy
\and
       Osservatorio Astronomico di Roma, Via Frascati 23,
       I-00040 Monteporzio Catone (Roma), Italy 
\and
       Affiliated to I.C.R.A.
}
\date{Received 9 June 2000; Accepted 13 July 2000 }

\maketitle

\markboth{X--rays from \src}{X--rays from \src}

\begin{abstract}

We report    the  results of a   BeppoSAX
observation   of the composite supernova remnant \src,  
located in the Galactic Center region. 
This new observation, in which the core of \src\  
is observed on--axis, confirms our previous results
obtained from a pointing in which the supernova remnant
 was serendipitously
observed at a large off--axis angle. 
The better quality of the new data allows us to confirm the 
non--thermal nature of the X--ray source coincident with the
central plerion, to significantly improve the estimate
of its spectral parameters, 
to put more stringent upper limits on the presence of
coherent pulsations (in the range of periods from 
4\,ms to $\sim$4$\times$10$^4$\,s), 
to obtain   marginal evidence
of the possible extended nature of the X--ray plerion
and to put a better upper limit on the 2--10 keV X--ray
emission from the 8$'$ diameter radio shell 
(L$_{\rm S}<3.4\times 10^{34}~{\rm d}_{10}^{2} $ erg s$^{-1}$). 
The X--ray spectrum is best fit
 with  a power--law model
with a photon index $\Gamma \sim$ 2, $N_{\rm H}\sim10^{23}$~\hcm 
and a 2--10 keV 
luminosity of $7.5\times 10^{34}~{\rm d}_{10}^{2} $ erg s$^{-1}$.
The broad band (radio--X--ray)  properties of the composite 
supernova remnant \src\ are indicative
of the likely presence of a young ($\sim$2,700 yr) neutron star 
in its central radio core.


\end{abstract}
 
\section{Introduction}

\begin{figure*}
\centerline{\psfig{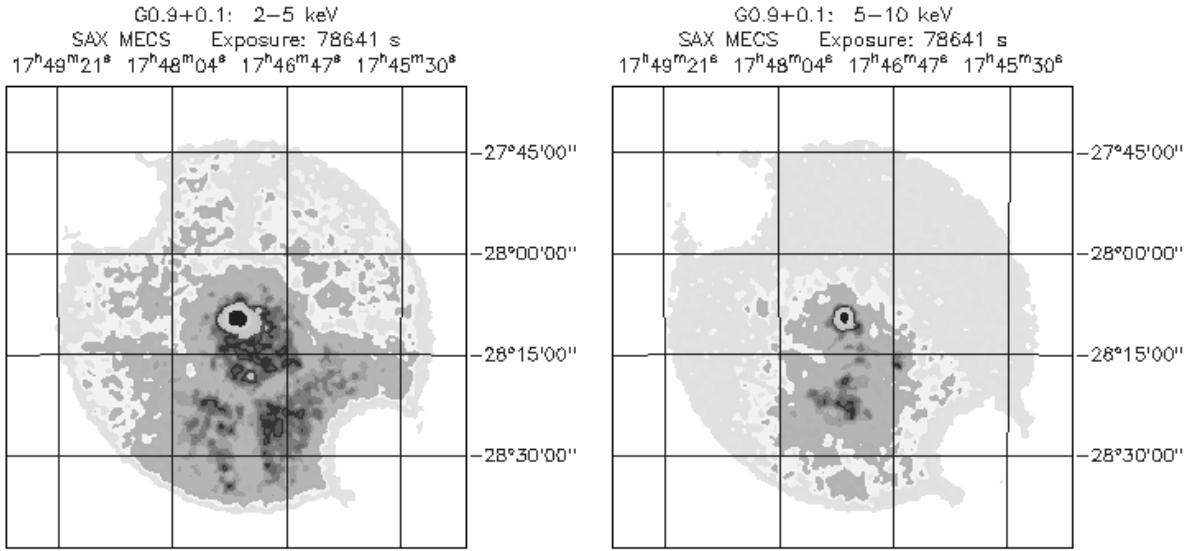}}
     \caption[]{BeppoSAX images of  \src\ obtained 
with the MECS instrument
in the 2--5~keV and 5--10~keV energy ranges 
(left and right panels, respectively).
Both  images have been smoothed with a
Gaussian with FWHM=1$'$.
Diffuse emission from the GC region and from the molecular
cloud Sgr~B2 is also visible in the Southern part of the field of view 
}
   \label{fig:mecs}
\end{figure*}
 

Supernova remnants (SNRs) are usually classified 
into three categories, based on their
radio morphology: plerions, shell--like and ``composite''.
About one tenth of   known  SNRs belongs to this 
last category (Green 1998). 
They are characterized by  a  radio/X--ray shell,  
related  to the expanding ejecta, plus a central synchrotron nebula, 
powered by the loss of rotational energy from a  neutron star
(Weiler  \& Sramek 1988).

The radio source \src\ was classified as an SNR by 
Kesteven (1968). Subsequent radio observations with the VLA  and the 
Molonglo Observatory Synthesis Telescope showed that G0.9+0.1 
can be classified as a  composite SNR. 
In fact, it  consists
of 
a radio polarized  core 
component  ($\sim$2$'$ diameter), surrounded by a  
shell ($\sim$8$'$ diameter) 
with a steeper radio spectrum 
(Helfand \& Becker  1987, hereafter HB87; Gray 1994). 
La Rosa et al. (2000)    recently  confirmed the composite 
morphology of the radio emission with VLA observations at  20  and 90 cm. 
They measured  spectral indices 
$\alpha$$_r$$\sim-0.77$ for the shell  and $\alpha$$_r$$\sim0.12$
for the core, typical of composite SNRs.

The position of \src\ is very close to the  Galactic Center (GC) 
direction,
and a distance estimate based on the $\Sigma$-D 
relation places it at $\sim$10 kpc
(Downes 1971).
Owing to the  high interstellar absorption ($N_{\rm H}\sim10^{23}$~\hcm), 
this SNR 
was not previously observed in  soft X--rays,
with the possible exception of a  marginal detection with the 
$Einstein$ Observatory (Helfand \& Becker 1987). 
During the BeppoSAX Survey of the GC region (Sidoli et al. 1999), 
we obtained the first clear X--ray detection of \src\ 
(Mereghetti et al. 1998). 
Here we report on a longer, follow-up BeppoSAX observation 
aimed
at better characterizing the properties of the 
X--ray emission from \src.

\section{Observations and data analysis}
 
The observations   were obtained with 
the   Medium Energy Concentrator Spectrometer  
(MECS, Boella et al. 1997) that covers the 1.8--10 keV energy range.
This is an   imaging detector that  provides  good spatial and  
energy resolution  
($\sim$8.5\% full width at half maximum (FWHM) at 6 keV)  over a circular
field of view with a diameter of $56'$. 
 
The location of G0.9+0.1 was observed  
on 1999 August 25--27  for a net exposure time of 78.6 ks 
in each of the 2   available MECS units.
The images in the soft (2--5 keV) and hard (5-10 keV)  energy ranges
are shown in   Fig.~\ref{fig:mecs}.
The central strong source,
located at R.A.$=17^{{\rm h}}~47^{{\rm m}}~22^{{\rm s}}$, 
Dec.$=-28\deg~09'~25''$ (J2000, 1$'$   uncertainty),
is coincident with the radio core of G0.9+0.1.
Due to the proximity of the GC,
significant diffuse X--ray emission is present in the whole
field of view (see e.g., Koyama et al. 1996;  Sidoli \& Mereghetti 1999).
This emission is particularly strong in the
  Southern part of the image, where the Sgr~B2 molecular
cloud  is located (Murakami et al. 1999; Sidoli 2000). 
 
Observations of the \src\ field 
carried out with the BeppoSAX 
high energy non--imaging  instrument (PDS, 15--200 keV)
are not useful in this context, because of the severe crowding
of contaminating hard X--ray sources in the GC region.

\subsection{Spectral analysis}

\begin{figure}
\vspace{0.5cm}
\centerline{\psfig{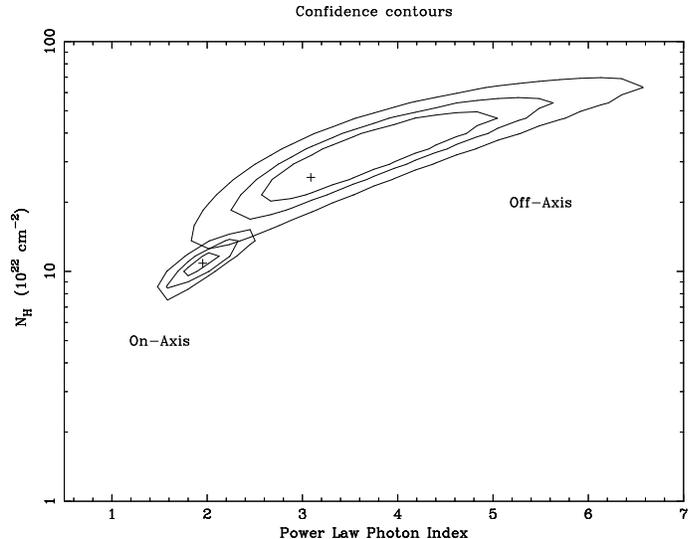}}
     \caption[]{Comparison of the spectral parameters (single power--law best
     fit model) derived for \src\  in the April 1997 off--axis
     observation (Mereghetti et al. 1998) 
and  the new on-axis one reported here. 68\%, 90\% and 99\% confidence level  contours are
displayed      
}
   \label{fig:oldnew}
\end{figure}

The location of \src\ is contaminated both by the 
GC thermal diffuse emission (kT$\sim10$~keV), 
and by  X--rays originating from the SgrB2 molecular cloud, with its
intense 6.4 keV iron line (see Fig.~\ref{fig:mecs}). 

To minimize the influence of these  components unrelated to \src, 
we  used a circular   extraction 
region of only 2$'$ in the spectral analysis.
In order to properly subtract the Sgr~B2 contribution,
the local background was estimated from the Southern half of an 
annular region centered on \src\  
with inner and outer radii of 5$'$ and $8'$. 
The resulting MECS  net count rate is 0.0247$\pm$0.0007 s$^{-1}$.
We rebinned the MECS counts in order to 
have at least 20 net counts in each energy bin, and to sample the
instrumental energy resolution with no more than 3 spectral channels.
 
We tried to fit the spectrum 
with   different     models: power--law,
thermal bremsstrahlung, black body, thermal hot plasma   ({\sc mekal}
in XSPEC). The results are reported in Table~\ref{tab:spe}.
The best fit model is a   power--law with
photon index $\Gamma=1.95$, absorption, $N_{\rm H}$, of $10^{23}$~\hcm, and 
flux F$ = 6.6\times 10^{-12}$ erg cm$^{-2}$ s$^{-1}$  (2--10 keV, 
corrected for   interstellar absorption). 
A thermal bremstrahlung gives an equally good fit, 
but the resulting high temperature (too high for a SNR)
means that it mimics a power--law in the energy range of our observations.

\begin{figure*}
\hspace{-4.5cm}
\centerline{\psfig{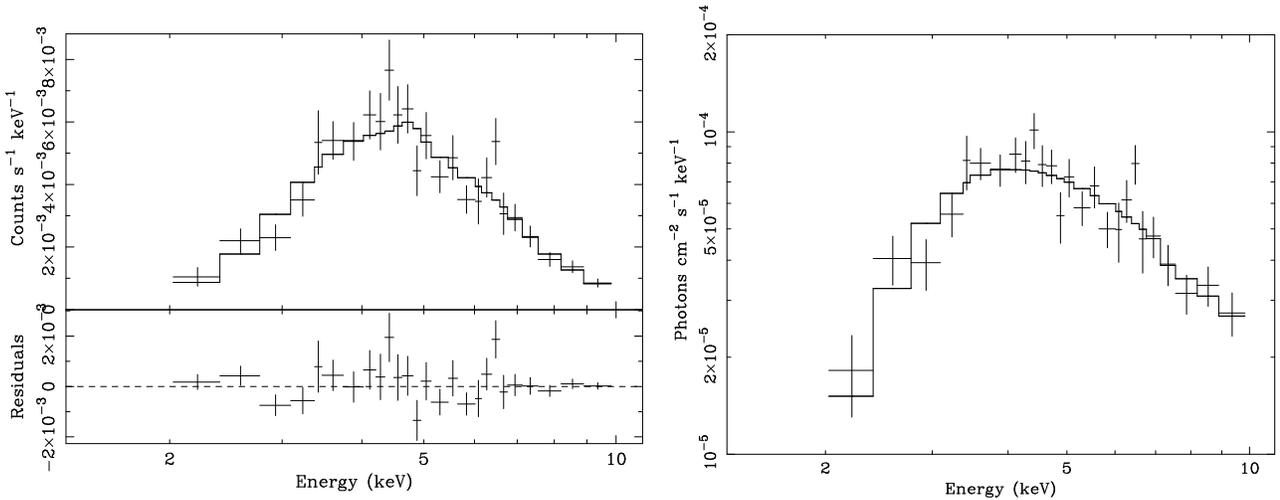}}
      \caption[]{Best-fit power--law model (the count spectrum  on the
      left, the photon spectrum on the right) 
      to the BeppoSAX spectrum of \src. 
See Table~\ref{tab:spe} for the parameters. The units of residuals are 
counts s$^{-1}$ keV$^{-1}$ 
}
   \label{fig:spe}
\end{figure*}

As shown in Fig.~\ref{fig:oldnew}, the power--law best
fit parameters are marginally   consistent
with those derived in our previous off--axis observation of \src, and their
uncertainties are much smaller. 
The spectrum is shown in Fig.~\ref{fig:spe}, together 
with the residuals. The addition of a narrow gaussian line in 
the range  6.4--7 keV
is not required by the data, yielding   a
normalization consistent with zero, within 90\% uncertainties.
The upper limit to the equivalent  width of a narrow Fe K line
is  330 eV, at a confidence level of 90\%.

Worse fits were obtained with a blackbody  
(see Table~\ref{tab:spe}), 
and with a  thermal plasma model ({\sc mekal} in XSPEC)
with abundance fixed at the solar value.   
Allowing the  abundance to vary,  
we obtain a  bremsstrahlung--like spectrum,
with abundance consistent with zero.
We also tried  to fit the spectrum with double--component models. 
However, no combinations of a power--law
plus a thermal model (blackbody or {\sc mekal}) 
gives significantly better results,
nor are   required by the   data.  

The 2--10 keV luminosity corresponding to the best fit power--law spectrum 
is L$_{\rm X} = 7.5\times 10^{34}$~erg~s$^{-1}$,
assuming  a  distance of 10 kpc.
  
\begin{table*}
\begin{center}
\caption[]{Results of the spectral fits. Uncertainties are quoted at 90\% 
confidence. All fluxes have been corrected for  interstellar absorption. 
}
\label{tab:spe}
\begin{tabular}{lllll}
\hline
\noalign {\smallskip}
Model          & Column density     & Parameter                          & $\chi^2$/dof  &Flux  (2--10 keV)\\
               &($10^{22}$ cm$^{-2}$)&                                    &           &($10^{-12}$ erg  cm$^{-2}$ s$^{-1}$ ) \\
\hline
Power law      & $10.9^{+2.4}_{-2.1}$        & $\Gamma=1.95^{+0.33}_{-0.30}$           & $28.2/23$ &     $6.6^{+1.3}_{-0.8}$      \\
Bremsstrahlung &$9.7^{+1.3}_{-1.4}$         &T$_{\rm br}=11^{+11}_{-3.5}$ keV    & $28.5/23$ &     $6.0^{+0.3}_{-0.5}$   \\
Black body     &$4.8^{+1.7}_{-1.4}$          &T$_{\rm bb}=1.75^{+0.17}_{-0.07}$ keV   & $34.4/23$ &     $4.1^{+0.5}_{-0.2}$   \\
{\sc mekal}  & $7.4^{+1.0}_{-1.0}$          & T$_{\rm M}=31^{+\infty}_{-13}$ keV   & $40.4/23$ &     $5.2\pm{0.3}$                  \\
\noalign {\smallskip}                       
\hline
\end{tabular}
\end{center}
\end{table*}

\subsection{Timing analysis}

MECS counts were extracted from a region with  
radius of 4\arcmin\   and used 
for the timing analysis, after the correction of their times of arrival 
to the solar system barycenter. 
No flux variations were seen during the   BeppoSAX observation. 
We searched for coherent pulsations over a 
wide range of periods (4\,ms -- $\sim$4$\times$10$^4$\,s),
both in the total energy range and   for the soft (2--5 keV) 
and hard (5--10 keV) bands separately,
without finding any significant signal.
In Table\,2 the 3$\sigma$ upper limits to the pulsed 
fraction, derived for the assumption of a sinusoidal modulation,
 are listed for selected periods.

\begin{table}
\begin{center}
\caption{ 3$\sigma$ upper limits on the pulsed fraction  
for a sinusoidal modulation 
(semiamplitude
of modulation divided by the mean source count rate)}
\begin{tabular}{lccc}
\hline 
Period & \multicolumn{3}{c}{Energy intervals} \\
 (s)   & \multicolumn{3}{c}{(keV)} \\  
 & 2--5 & 5--10& 2--10 \\   
\hline 
10000  &  22\% & 33\% & 18\%   \\
1000   &  18\% & 28\% & 16\%   \\
100--1 &  20\% & 28\% & 16\%   \\
0.1    &  40\% & 43\% & 30\%    \\
0.01   &  41\% & 49\% & 39\%    \\
0.004  &  50\% & 68\% & 38\%    \\
\hline 
\end{tabular}  
\end{center}
\noindent 
\end{table}

\subsection {Radial profile}

In order to check whether the X--ray emission originates from
an extended source or from an unresolved point source, 
we examined the radial distribution of the \src\ counts.
We used as a comparison a 25 ks  observation of  Cyg X--1  
obtained  in 1998 October. 
We selected only time intervals in which the BeppoSAX
attitude was determined with  two star trackers 
(in fact small deviations 
may be present 
when no star tracker is active or when only one is in use). 
This procedure restricted us to consider about 80\% of the on--source time.
We then  extracted the  counts 
from 6 equally spaced  Northern semi--annular  
regions within  3$'$ from the centroid of \src.
The Southern sector, containing the contaminating Sgr~B2 emission, 
was excluded from the analysis.
For the \src\ background determination we used the Northern semi--annular
region of 5$'$--8$'$. 
The same procedure was used for the reference observation of Cyg X--1.

\begin{figure}
\centerline{\psfig{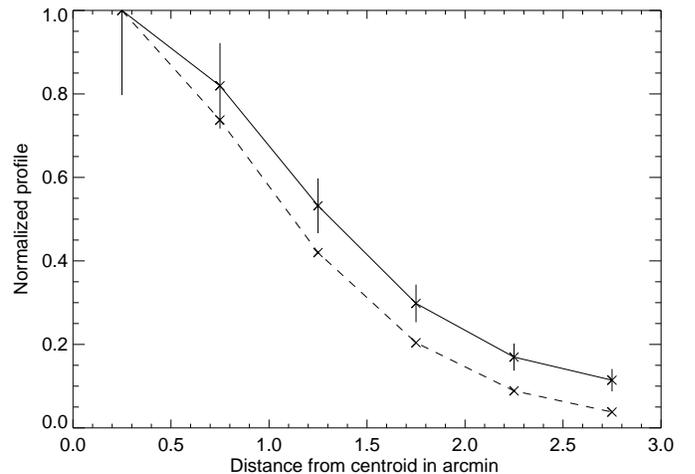}}
     \caption[]{Comparison of the \src\ radial profile  (solid line)
in the 2--10 keV energy range, with the radial profile 
of Cyg X--1 (dashed line)  taken as reference for the MECS point
spread function (PSF). The  profiles are  normalized at the peak.  
From the comparison of the two profiles there is   indication of  
an extended nature of the \src\ plerionic emission 
at distances from the peak further than $\sim$1$'$.  
However, this could 
be also due to a 
still remaining contamination from the intense GC background  
}
   \label{fig:psf}
\end{figure}

The  resulting radial profiles, normalized at the peak, 
are plotted  in Fig.~\ref{fig:psf}.  
There is evidence that the profile of \src\ is slightly broader
than that expected for a single point--like source. 
This might be due to the contribution of a central 
extended 
synchrotron nebula,
although we cannot exclude that the difference in the
two radial profiles is due to the presence of an underlying diffuse emission,
not properly subtracted by our background estimate.


\subsection{Limits on  shell X--ray emission}
\label{sec:shell}

To estimate an upper limit to the X--ray emission from the radio 
shell of \src\ (located   4$'$ from the radio core) 
we considered the counts
from an annular region from 3$'$ to 5$'$ from the centroid
of the emission. 
Fitting the resulting spectrum with an absorbed {\sc mekal} model
(N$_{\rm H}$ fixed at 10$^{23}$~cm$^{-2}$) we derived  a 2--10 keV
flux  F$_{\rm s}<3$$\times10^{-12}$~erg~s$^{-1}$~cm$^{-2}$
(corrected for   interstellar absorption), that
translates into a luminosity of $\sim3.4\times10^{34}$~erg~s$^{-1}$.
Of course, this represent an upper limit to the shell luminosity, since
some of the counts could be due to Sgr~B2.

\begin{figure}
\centerline{\psfig{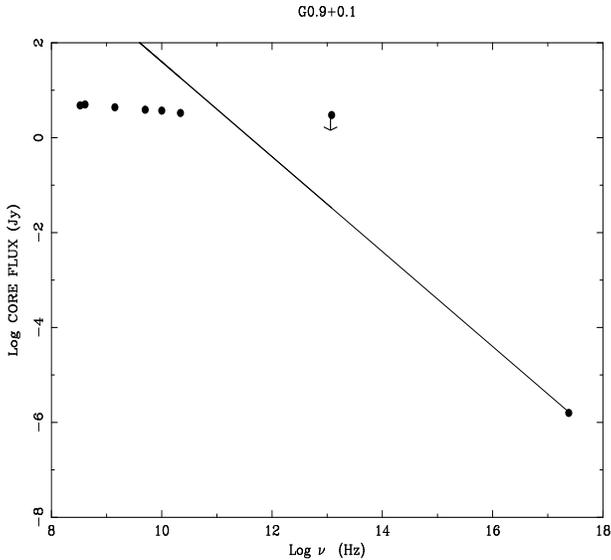}}
      \caption[]{\src\ broad band spectrum (adapted from HB87): 
radio data are from 
HB87 and La Rosa et al. (2000), the infrared upper limit is from HB87 and
the X--ray flux density at 1~keV is from our new BeppoSAX observation.
The straight line indicates the extrapolation of
the slope of the X--ray spectrum 
back to the radio range
}
   \label{fig:broad}
\end{figure}

\begin{figure*}
\centerline{\psfig{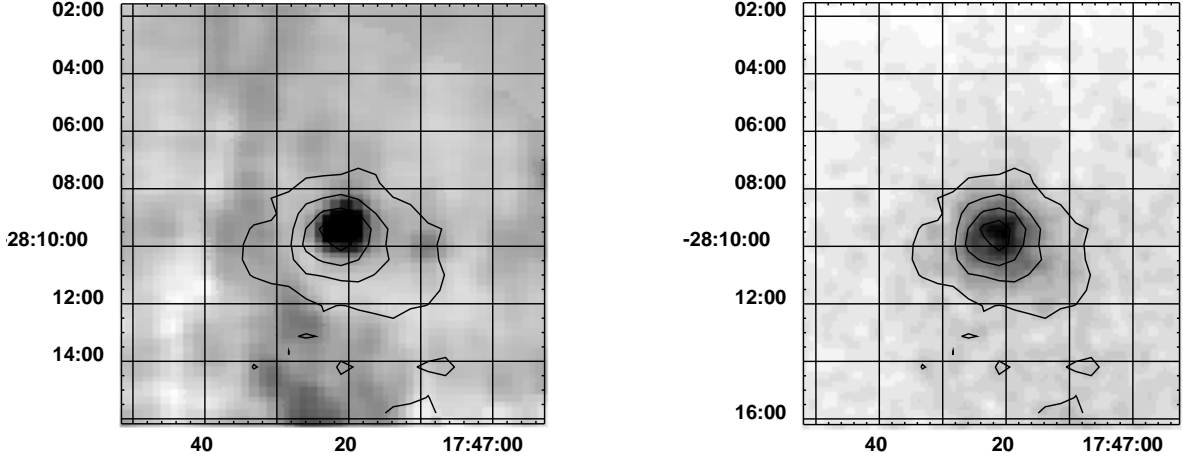}}
     \caption[]{Radio and X--ray (2--10 keV) images of \src\ in the same 
     coordinate system (J2000): on the left, the 
NRAO VLA Sky Survey (NVSS) image at 1.4 GHz with the 
X--ray contours (MECS) overlayed;  part of the   radio
shell is also visible on the left, at about 4$'$ from the radio core. 
On the right, the central 
part of the MECS image with  80\%, 60\%, 40\% and 20\% 
of the X--ray peak value contours overlayed. 
From the comparison
it is evident that the X--rays originate from the radio core and not from the
shell  
}
   \label{fig:radiox}
\end{figure*}
 

\section{Discussion}

The  new MECS observation presented here, thanks to  the better on--axis
spatial and spectral resolution and to the better
statistics, allows us  to confirm  both the association of the X--rays
with the central radio core of \src\ and their non-thermal  nature. 

\subsection{Radio and X--rays}

A comparison of the X--ray and radio morphology of \src\ is presented in
Fig.~\ref{fig:radiox}, where it is clearly evident
that the observed  X--rays originate from the plerionic component rather than 
from the shell. 
These radio data are from   the 
NRAO VLA Sky Survey (NVSS,  Condon et al. 1998) provided by
the public archive available 
at {\em http://skyview.gsfc.nasa.gov}.
We note that \src\ displays a different morphology 
depending on the radio frequency: 
while the image at  1.4 GHz reported here shows  
the plerion and  the Eastern  part of the shell (located
at a distance of 4$'$ from the radio core),   at 90~cm 
(HB87; La Rosa et al. 2000) the shell is better defined and the shell 
emission is stronger in its Western rim.  
 
The upper limit   derived
in Sect.~\ref{sec:shell}, can be compared to the  
shell X--ray luminosities of other
composite supernova remnants reported by HB97. 
After converting to the
0.5--4.5 energy range, we obtain 
L$_{\rm Xshell}$ $<8\times10^{34}$~erg~s$^{-1}$. 
From our new X--ray data (extrapolated to the 0.5--4.5  keV
range) and the radio observations (HB87;
L$_{\rm Rcore}$=7$\times10^{34}$~erg~s$^{-1}$), 
we find the following ratios:
L$_{\rm Xc}$/L$_{\rm Xs}$$>$1.2, L$_{\rm Xc}$/L$_{\rm Rc}$=$1.4$,
where ``c'' and ``s'' indicate the core and shell components.
These values place \src\ well inside the ranges observed in other
composite supernova remnants (see  e.g., the correlations reported by HB87).

\subsection{Energetics and age}

As shown in Fig.~\ref{fig:oldnew} the spectral parameters 
of the central component
are   consistent with those reported 
earlier (Mereghetti et al. 1998)
and there is no evidence for flux variations between 1999 
August   and  1997 April.
This is consistent with the hypothesis that the X--ray emission derives
from a synchrotron nebula powered by an energetic pulsar at the center
of \src. 
It is also likely that part of the emission is due to the pulsar itself.
The upper limits derived in our periodicity search do not allow us
to exclude the presence of a pulsar with a relatively large
pulsed fraction (below $\sim$30\%--40\%), 
in particular for periods shorter than 0.1\,s.

The  power--law  photon index, $\Gamma\sim$1.95, is 
in agreement with the value of 2 typical 
of other well studied X--ray synchrotron  nebulae 
(Asaoka \& Koyama 1990)
powered by young energetic pulsars. 

Seward \& Wang (1988) derived a  relation between the
X--ray luminosity and rotational energy loss of neutron stars
powering synchrotron nebulae. 
Although this is just an   empirical law, subject to 
large uncertainties, it can be used to roughly estimate   the 
possible parameters of the neutron star in \src.
Assuming I=10$^{45}$ g~cm$^{2}$ for the neutron star moment of inertia,
the Seward \& Wang (1988) relations for the pulsar period,
period derivative and the surface magnetic field are:

\begin{equation}
{\rm P }= 0.25[{\rm t}(10^{3} {\rm yr})]^{-1/2} [{\rm \dot E} (10^{37} {\rm erg \,
s}^{-1})]^{-1/2}  \, {\rm s}
\end{equation}

\begin{equation}
{\rm \dot P }= 1.58\times 10^{-11}[{\rm P}({\rm s})] [{\rm t}(10^{3} {\rm yr})]^{-1} \,
{\rm s \, s}^{-1}
\end{equation}

\begin{equation}
{\rm B_{o} }= [{\rm P(s) \dot P} (10^{-15} {\rm s \, s}^{-1})]^{1/2}\times
10^{12}\, {\rm G}
\end{equation}

\noindent The estimated age for the remnant, as indicated by 
the diameter of the radio shell ($\sim$12 pc at the
GC distance)  is between 
$\sim$1,100~yr (free expansion with  velocity v$\sim10^{4}$~km~s$^{-1}$)
and $\sim$6,800~yr (adiabatic expansion; Mereghetti et al. 1998).
Thus we obtain  P=0.19~s, ${\rm \dot P}=2.8\times 10^{-12}$~s$\,$s$^{-1}$,
and a magnetic field B$_{\rm o}\sim$$23\times 10^{12}$~G for the free
expansion age  or 
P=0.078~s, ${\rm \dot P}=1.8\times 10^{-13}$~s$\,$s$^{-1}$,
and a surface magnetic field B$_{\rm o}\sim$$3.8\times10^{12}$~G 
for the adiabatic age
(${\rm \dot E}=1.5\times10^{37}$  erg~s$^{-1}$ from our X--ray data).
 
The age of the plerion can also be estimated
from the frequency break in the broad band spectrum
from radio to X-rays 
shown in Fig.~\ref{fig:broad}.
Extrapolating the slope of the 
now well defined 
BeppoSAX spectrum back to the radio range
and assuming that
a single spectral break occurs and that the spectrum
is straight after the break, we can place the frequency
of the break $\nu_b$ at $\sim$10$^{11}$ Hz.
The energy in relativistic electrons is:

\begin{equation}
{\rm E_e} = 4\times10^{47} \Bigg(\frac{B}{10^{-4} \, {\rm G}}\Bigg)^{-3/2}
\Bigg(\frac{\nu_b}{10^{11} \,  {\rm Hz}}\Bigg)^{1/2}  \, {\rm ergs,}
\end{equation}
\label{eq:bnebular}
where B is the nebular magnetic field.
Assuming equipartition between particles and magnetic fields
(U$_e$=(4/3)U$_B$=B$^{2}$/6$\pi$), B can now be calculated,
considering a core component volume of $\sim3\times10^{57}$~cm$^{3}$ (HB87)
and $\nu_b$=10$^{11}$ Hz. This gives B=6.7$\times10^{-5}$~G.
This translates into
a total energy for the magnetic field and the relativistic
particles of E$\sim1.3\times10^{48}$~erg.
So, an independent estimate of the age of the remnant is
t \,= \, (E$/ \dot {\rm E})\, \approx$ \, 2,700~yr,
closer to the free-expansion approximation, than to the adiabatic age.

\subsection{Other SNRs in the GC region}

The location of the composite SNR \src\ in the GC  invites
a comparison with other SNRs located in this same region.
The Green Catalogue (Green 1998)
 lists seven SNRs in the direction of the  GC 
 ($|l|<1.0$$\times|b|<1.5$), six shell--like 
 and only one  composite (\src).
Besides    \src, 
only two of them have been reported to clearly emit X--rays: 
G359.1--0.5, from its central region surrounded by the radio shell,
 and G359.0--0.9, from a part of its radio shell (Bamba et al. 2000; ASCA).
While G359.1--0.5 seems to be located at the GC distance, 
 G359.0--0.9 is probably  a foreground object.
 
Another possible  SNR,  the 
SgrA~East halo (Pedlar et al. 1989),
has been recently
reported to emit thermal X--rays with kT$\sim$0.6 keV 
(Sidoli \& Mereghetti 1999). 
The SgrA East halo is a  radio source with a diameter of $\sim$20 pc 
located near the center of our Galaxy. It is a distinct object from
the nearby   SgrA~East  (Pedlar et al. 1989) 
and is not included in the Green catalogue.

The  lack of X--ray detections from the other SNRs  
in the GC region could be due to
several reasons: 
their shells may  not emit X--rays,   
due to an evolved age of the remnants 
(in the case of thermal emission) 
or, as suggested by Bamba et al. (2000),
 because of  a reduced lifetime   of the electrons emitting  synchrotron
X--rays caused by a higher magnetic field in the GC region (in the case
of non--thermal emission from the shell). 
In addition,  
if  their shells  emit soft X--rays, 
their  observability is 
hampered  by the severe interstellar absorption (especially below 2 keV) 
and by the presence of contaminating X--ray sources,  
both point--like and diffuse. 
It is thus not surprising that definite detections
have been obtained 
when the emission is  hard (as in the case of the plerion \src),
or particularly bright   (as in the case of the SgrA~East halo). 

SNRs are thought to contribute to the diffuse X--ray emission
observed from the GC, especially at soft X--rays.
However, the number of SNRs known to emit high energy radiation is 
too low to account for this emission (e.g. Koyama et al. 1996), even if
the reason for this discrepancy could only be  an  observational
bias.

\section{Conclusions}

The new BeppoSAX data reported here,  the first X--ray observation
pointed at the center  
of the composite SNR \src,
have confirmed the non--thermal origin of the X--ray
emission that was suggested by the previous
serendipitous observation
(Mereghetti et al. 1998; Sidoli et al. 1999).
Such data could not  
completely rule out a thermal origin of the X--ray emission, due to the
poor statistic of the off--axis spectrum.
This follow-up   observation has allowed us
to significantly  better
constrain the spectral parameters and to safely conclude
that the X--rays observed from \src\ are due to non--thermal emission 
from its  radio core. 

A thermal model, either originating from  the surface of a   
neutron star, or  from a hot diffuse plasma filling the 
central part of the  SNR, can now be excluded 
 as the main contributor to the observed X--rays.
A double component model, non--thermal 
plus thermal (blackbody or {\sc mekal})
emission components, is not required by the BeppoSAX data.

There are some indications of a diffuse nature of the X--ray emission
from \src. However, 
since a strong contamination from the local background
exists, 
XMM--Newton and $Chandra$ observations, with their superior sensitivity and 
spatial and spectral resolution,  are needed 
to confirm this finding and to 
discriminate between a point source 
(i.e. a neutron star) and a synchrotron nebula 
and to possibly detect
X--rays from the radio shell.

\begin{acknowledgements}
The \sax\ satellite is a joint Italian-Dutch programme. 
L. Sidoli and F. Bocchino acknowledge an ESA Fellowship. 
We thank A. Parmar for helpful discussions.
This research has made use of  NVSS radio data obtained through 
$SkyView$, provided 
by the NASA/Goddard Space Flight Center.
\end{acknowledgements}

\end{document}